\documentclass{article}

\usepackage[colorlinks,urlcolor=blue,linkcolor=blue,citecolor=blue]{hyperref}
\usepackage[subrefformat=parens,labelformat=parens]{subcaption} 
\usepackage{color,array}
\usepackage{graphicx}
\usepackage{cite}
\usepackage{amsmath,amssymb,amsfonts}
\usepackage{tabularx} 
\usepackage{booktabs}
\usepackage{multirow}
\usepackage{multicol}
\usepackage[table]{xcolor}
\usepackage[margin=1in]{geometry}
\newcolumntype{M}[1]{>{\centering\arraybackslash}m{#1}}
\newcolumntype{J}[1]{>{\arraybackslash}m{#1}}

\definecolor{level1}{HTML}{AED6F1}
\definecolor{level2}{HTML}{5DADE2}
\definecolor{level3}{HTML}{3498DB}
\definecolor{myblue}{HTML}{115CE7}

\newtheorem{definition}{Definition}

\title{Cooperative Resilience in Artificial Intelligence
Multiagent Systems}
\author{
    \small Manuela Chacon-Chamorro \and
    \small Luis Felipe Giraldo \and
    \small Nicanor Quijano  \and
    \small Vicente Vargas-Panesso  \and
    \small César González  \and
    \small Juan Sebastián Pinzón \and
    \small Rubén Manrique \and
    \small Manuel Ríos 
    \and
    \small Yesid Fonseca  \and
    \small Daniel Gómez-Barrera
    \and
    \small Mónica Perdomo-Pérez \thanks{This work was supported by Google through the Google Research Scholar program and the UniAndes-DeepMind Scholarship 2023.\\
    M. Chacon-Chamorro (\mbox{m.chaconc}),  L.F. Giraldo (\mbox{lf.giraldo404}), 
N. Quijano (\mbox{nquijano}),  V. Vargas-Panesso (\mbox{jv.vargas}), 
C. González (\mbox{cl.gonzalezg}), J.S. Pinzón (\mbox{js.pinzonr}), 
R. Manrique (\mbox{rf.manrique}),  Y. Fonseca (\mbox{y.fonseca}) and D. Gómez-Barrera (df.gomezb) are with the Universidad de los Andes, Colombia \mbox{[@uniandes.edu.co]}.\\
M. Ríos (manrios) is with Center of Excellence in Analytics and Artificial Intelligence Bancolombia [@bancolombia.com.co].\\
M. Perdomo-Pérez (tatiana.perdomo) is with the Universidad de Ibagué, Colombia [@unibague.edu.co].}
}

\date{\small \today}

\begin{document}

\maketitle
\hrule
\begin{abstract}
Resilience refers to the ability of systems to withstand, adapt to, and recover from disruptive events. While studies on resilience have attracted significant attention across various research domains, the precise definition of this concept within the field of cooperative artificial intelligence remains unclear. This paper addresses this gap by proposing a clear definition of `cooperative resilience' and outlining a methodology for its quantitative measurement. The methodology is validated in an environment with RL-based and LLM-augmented autonomous agents, subjected to environmental changes and the introduction of agents with unsustainable behaviors. These events are parameterized to create various scenarios for measuring cooperative resilience. The results highlight the crucial role of resilience metrics in analyzing how the collective system prepares for, resists, recovers from, sustains well-being, and transforms in the face of disruptions. These findings provide foundational insights into the definition, measurement, and preliminary analysis of cooperative resilience, offering significant implications for the broader field of AI. Moreover, the methodology and metrics developed here can be adapted to a wide range of AI applications, enhancing the reliability and effectiveness of AI in dynamic and unpredictable environments.

\end{abstract}

\vspace{0.2 cm}
\hrule

\section{Introduction}

\label{sec:introduction}

Understanding how systems withstand and adapt to adversity has become a focal point for researchers. This capability is referred to as resilience. The concept of resilience has been extensively explored across various domains, ranging from game theory \cite{1_1, 1_2, 1_4, 1_6}, artificial intelligent (AI) systems \cite{2_1,2_2,2_3,2_4,2_5}, engineering \cite{3_1,3_4,zambrano2021you}, psychology \cite{4_1, 4_2, 4_3,4_4,5_1,4_10}, economy \cite{6_1,6_2,6_3}, social science \cite{7_1,7_2}, network science \cite{8_1, 8_2, 8_3}, dynamical systems theory \cite{10_1,10_2}, and ecology \cite{11_1,11_2,11_3}. Particular interest lies in understanding how systems that involve collective action, whether from humans, machines, or both, exhibit resilience as an emergent property from their interactions. These systems are encompassed within the domain of cooperative AI. 

Cooperative AI systems operate in complex, dynamic environments \cite{dafoe2020open}. Interactions with various actors, whether human or machine, add further complexity and uncertainty. This makes them more susceptible to disruptions and failures, as they must continually adapt to changes while maintaining efficient responses \cite{2_2, 3_4, 8_2}. Consequently, understanding and transferring the concept of resilience from other domains to cooperative AI systems is crucial. Doing so can inspire the development robust AI architectures and methodologies \cite{han2023synergistic}. Emphasizing resilience ensures that these systems remain adaptable and persistent in the face of disruptions. 

While the concept of resilience has been studied extensively across various domains, its definition in the context of cooperative AI problems remains unclear. This situation represents a need not only to contribute to unifying terminology in this field but also to understand how the capacity to withstand adversity emerges in systems of this nature. Furthermore, to characterize these systems and enhance their resilience capacity, it is necessary to establish a method for quantifying this property. This method should provide a measurable value that must be related to the definition of the concept.

Several indicators have been proposed to quantify resilience. For instance, in critical infrastructure \cite{ayyub2014metrics, wang2023metrics, argyroudis2022metrics, gerges2022metricsIndex}, or in community systems \cite{serfilippi2018metricsIndex, cimellaro2016peoples}. Through a review of such metrics, two general approaches for quantifying resilience emerge. The first approach involves analyzing time-dependent measures to establish the system's performance and compare it against its performance during disruptive events. The second approach involves resilience indices, using instantaneous measures of dimensions linked to the system's performance. These measures can be estimated before, during, or after the disruptive event and are used to assess changes in the system's performance \cite{serfilippi2018metricsIndex, gerges2022metricsIndex}. While the proposed measures aim to quantify resilience, they often lack guidance on applying this concept across various domains. This gap is particularly challenging in environments where collective interactions between humans, machines, or both are crucial to cooperative AI.

To fill this gap, the first contribution of this paper is to define the concept of resilience in cooperative AI. This definition is consistent with the work in \cite{dafoe2020open}, which unified concepts in cooperative AI. As we explain later in the paper, each element of this definition provides an important aspect of the concept of cooperative resilience. Also, we introduce a methodology to quantify cooperative resilience that aligns with the proposed definition, establishing a series of characteristics, assumptions, and stages designed to quantify cooperative resilience. The proposed methodology is crafted to be adaptable across various contexts, aiming to characterize resilience within cooperative AI systems. This approach will aid in understanding the emergence of resilience and guide strategies for enhancing it.

The proposed methodology is then validated through experiments conducted in the field of cooperative AI. The experiments are developed in Melting Pot 2.0 \cite{agapiou2022melting}, a multiagent domain for studying scenarios in which social dilemmas can arise. In particular, we study the Common Harvest Open scenario. In this scenario, agents interact with a resource patch of apple trees that regenerate based on current availability. This configures a social dilemma: if all apples are consumed, no more will grow, so agents must have a social understanding of their actions to avoid overharvesting and ensure the resource's sustainability. The experiments utilize agents based on Reinforcement Learning (RL) and Large Language Models (LLM) to explore this dilemma. Two disruptive events are tested: environmental changes and the introduction of agents with unsustainable behaviors. This scenario is particularly useful for evaluating cooperative resilience because it highlights how agents' collective well-being and the system's ability to handle disruptions are interrelated. The experimental results show that the proposed metric highlights aspects of agents' ability to resist, adapt, and transform in the face of disruptions, which other metrics may overlook.

This document is organized as follows: in Section \ref{sec:definiton} the definition of cooperative resilience and the analysis of key elements is presented. Section \ref{sec:methodology} introduces the proposed methodology for measuring cooperative resilience. Section \ref{sec:cases} elaborates on specific examples applying the suggested methodology. These three sections aim to establish an understanding of the cooperative resilience concept and provide a framework for measurement aligned with the definition. Finally, conclusions and further discussions are presented in Section \ref{sec:conclusions}.

\section{Defining Cooperative Resilience}

\label{sec:definiton}

\begin{figure*}[ht]
\centerline{ \includegraphics[width = 0.7\textwidth]{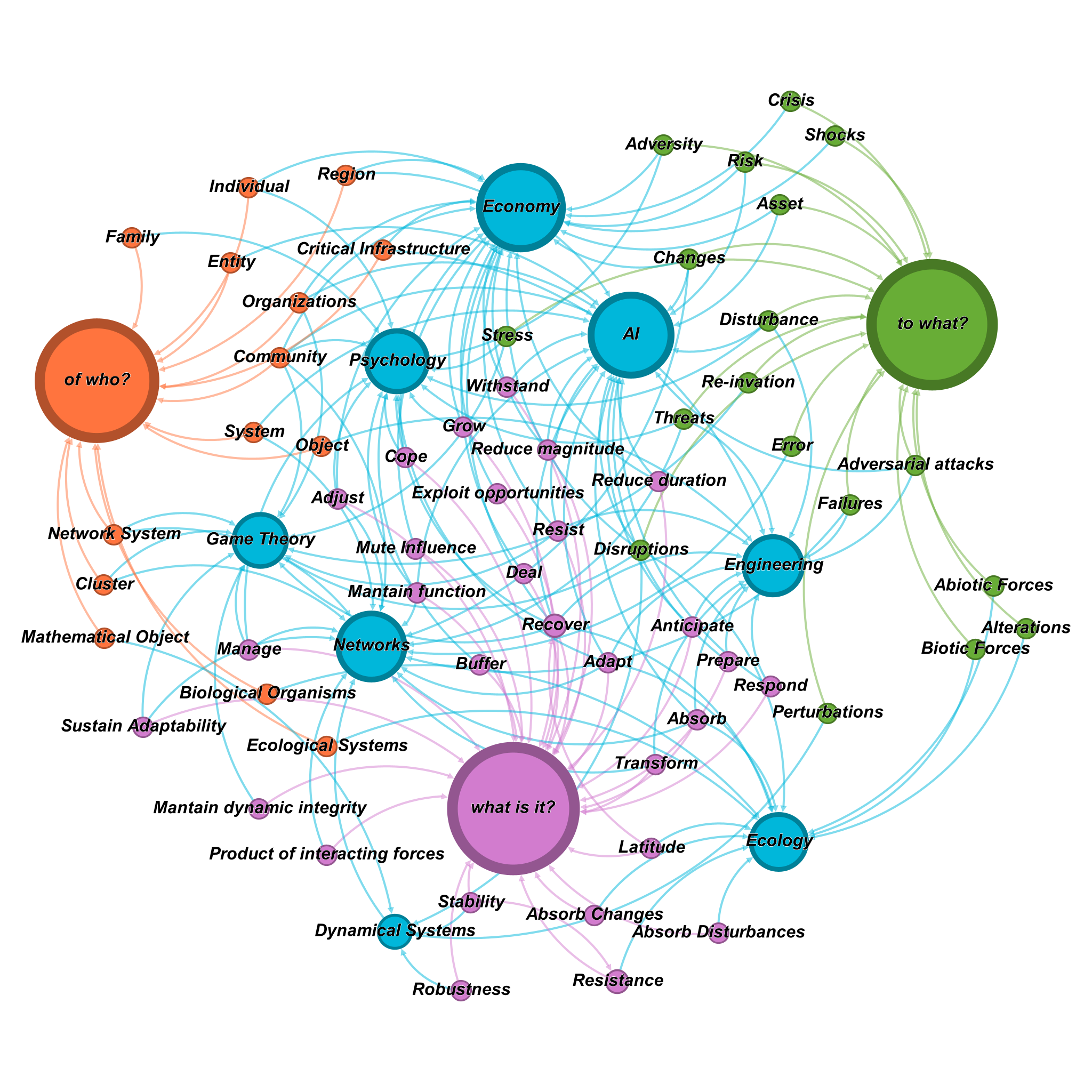}}
\caption{Keyword map of resilience across diverse fields and contexts, addressing guiding questions.}
\label{fig:KeyWordsImage}
\end{figure*}

Several key factors from definitions of resilience across various disciplines contribute to defining cooperative resilience. The essence of defining resilience lies in identifying the resilient entity (who?), the actions that define resilience (what is it?), and recognizing the disruptive event (to what?). These key questions, along with their corresponding keywords, are illustrated in Fig. \ref{fig:KeyWordsImage}. This figure summarizes a review of resilience concepts across various fields and demonstrates the broad scope of the concept across multiple disciplines\footnote{See the detailed review in the supplementary file.}. In the figure, fields are represented as blue nodes, while the guiding questions are shown as nodes in orange (who?), green (to what?), and purple (what is it?). The edges illustrate the relationships between the fields and these questions, as well as the interdisciplinary connections of the concept.

For instance, in ecology, resilience is related to verbs such as absorb, transform, and respond \cite{11_2}. It encompasses elements such as resistance and latitude, representing the extent to which a system can be altered before losing its capacity to recover, as well as stability \cite{11_1, 11_3}. Here, the resilient entity is an ecological system, and the disruptive event involves disruptions in population dynamics \cite{11_2}. In engineering, these disruptions often involve failures \cite{9-1, 9-2}, errors, or adversarial attacks \cite{2_2, 2_3}, with resilience linked to terms like resist, recover, and adapt \cite{3_1, 3_2, 3_3, 3_4}. In psychology, a resilience entity spans from individuals to groups such as families and communities \cite{4_1,4_2,4_3,4_4, 4_5, 4_6}, with disruptions related with stress, threats, and disturbances that are embedded in life events \cite{4_1, 4_2, 4_3, 4_4}. In economics, resilience is linked with actions such as withstand, grow, or resist and the disruptions are associated with terms like risk, crisis, and change, and  \cite{6_1, 6_2, 6_3}. Resilience in dynamic systems pertains to how these systems respond to disturbances that can include external factors, changes in initial conditions, or variations in parameters \cite{10_1, 10_2}. In network science, the resilient entity could be clusters of interacting agents responding to disturbances, and the disruptions can include failures, errors, threats, or changes in their environment \cite{8_1, 8_2, 8_3}.

In each field of study, resilience is defined by a resilient entity, a disruptive event affecting its normal behavior, and verbs describing the entity's actions before, during, and/or after the disruption. Based on a literature review covering the concept of resilience across diverse fields, emphasizing the previously mentioned key elements and considering the scope and terminological unification in cooperative AI \cite{dafoe2020open}, we introduce a novel concept aligned with cooperation paradigms: `cooperative resilience,' proposed in Definition \ref{definition:cooperativeResilience}.

\begin{definition}{}
 Cooperative resilience is the ability of a system, involving the collective action of individuals ---whether humans, machines, or both--- to anticipate, prepare for, resist, recover from, and transform in the face of disruptive events that threaten their joint welfare.
 \label{definition:cooperativeResilience}
\end{definition}

In Definition \ref{definition:cooperativeResilience}, the resilience entity is identified as a system comprising a collective of individuals, whether humans or machines, interacting with each other. This definition incorporates five key actions: \textbf{anticipate}, \textbf{prepare}, \textbf{resist}, \textbf{recover}, and \textbf{transform}. These verbs represent critical moments that span from the pre-disruptive event stage to its subsequent management. By including these verbs, resilience is analyzed not only as an inherent system capability but also as a \textbf{process} composed of a series of fundamental stages. In particular, the verbs `anticipate' and `prepare' are related to the static capabilities of resilience, often referred to as capitals in the literature \cite{serfilippi2018metricsIndex}.

Additionally, resilience also encompasses a reinforcement effect, wherein disruptive events experienced by a system could lead to learning about how to react, act, and prepare for future occurrences. The ability to transform is included as part of promoting either a positive, neutral, or negative change in the system's performance and how it attains a different configuration. These aspects are inspired in psychology and economics, where systems exhibit resilience when strategies are developed to confront disruptive events and foster growth. The notion of growth is relevant in defining resilience, often referred to as the capacity to exploit opportunities.

The last part in Definition \ref{definition:cooperativeResilience} focuses on the aspect of the expected behavior of the system in absence of disruptive events. This factor is crucial for measuring and interpreting resilience in any context. To assess how resilient a system is, it is necessary not only to characterize its prior configuration, but also to approach the concept from the dynamics of the system's performance and how this expected behavior should manifest. It is additionally emphasized that disruptive events pose a risk to the collective well-being of the system, implying that the expected behavior will be addressed in terms of joint welfare. This consideration is introduced with the recognition that ``AI research is aimed at helping individuals, both humans and machines, find ways to enhance their joint welfare,''\cite{dafoe2020open} as highlighted in the cooperative AI approach.

Also, Definition \ref{definition:cooperativeResilience} specifies that disruptive events pose a risk, emphasizing the stochastic nature inherent in these occurrences. In resilience literature, a disruptive event is identified as a phenomenon that could be external, internal, or even an attack on the resilient entity, disturbing its normal operational conditions. Given the entity's focus, the stochastic nature of disruptive events lies in their randomness and unpredictability in terms of timing or magnitude.

\section{Measuring Cooperative Resilience}
\label{sec:methodology}

Once the definition of Cooperative Resilience has been proposed, it is essential to establish a consistent measurement approach. This section proposes a methodology that comprehensively captures all aspects of the concept. It is important to note that resilience, as outlined in Definition \ref{definition:cooperativeResilience}, depends on the random nature of disruptive events. This randomness is characterized by the probability of occurrence ($p_s$) and the probabilistic magnitude of impact ($v_s$). These parameters are related to risk and vulnerability and could be associated with various sources depending on the system. Therefore, the measure is determined by the realization of the random disruptive event, and this realization, based on specific values of $p_s$ and $v_s$, is referred to as a scenario.

\begin{figure}[htbp]
\centerline{ \includegraphics[width = 0.7\textwidth]{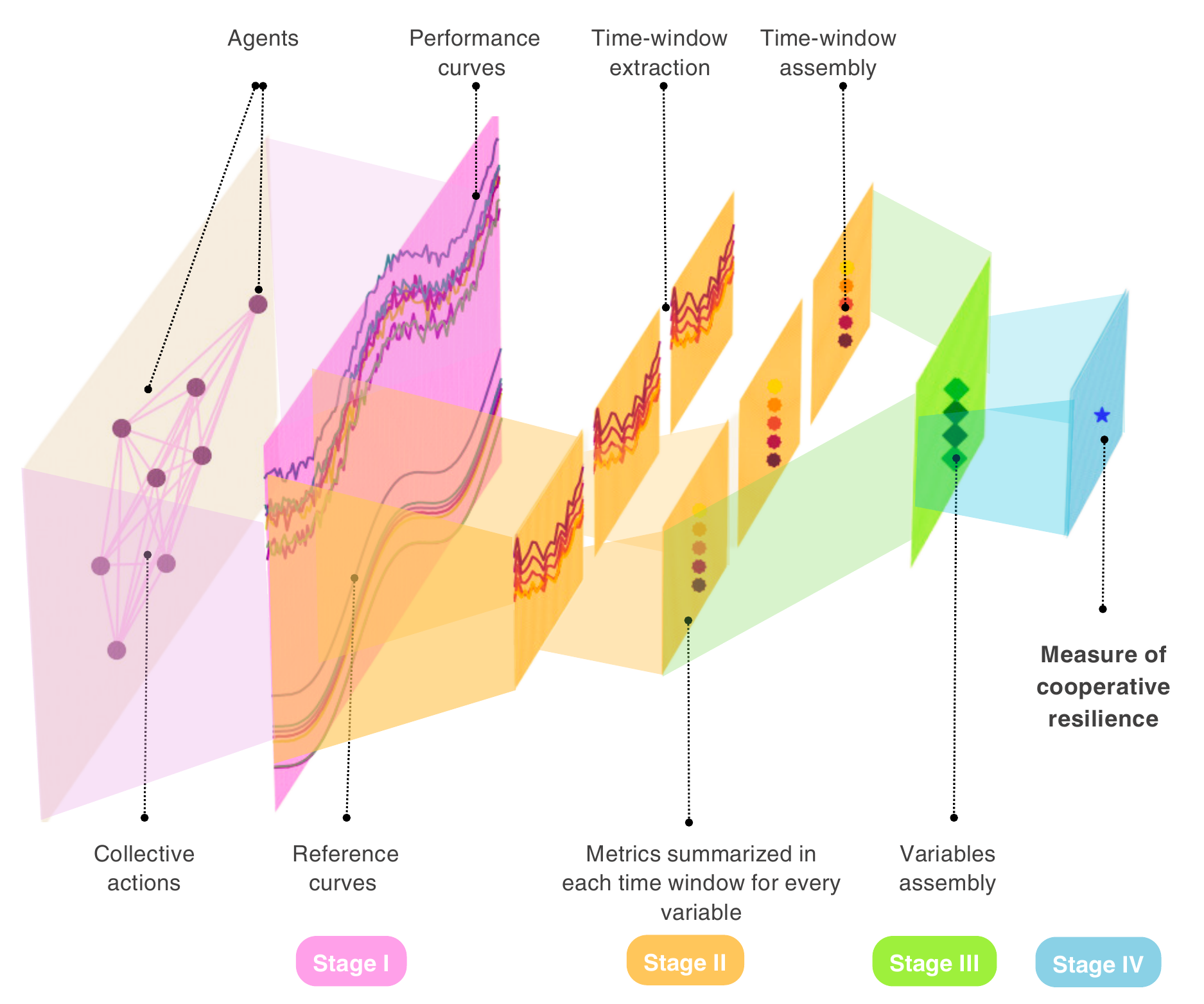}}
\caption{Diagram illustrating the proposed methodology for measuring cooperative resilience.}
\label{fig:methodologyResilience}
\end{figure}

The proposed methodology consists of four stages summarized in Fig. \ref{fig:methodologyResilience}. In the first stage, we begin by assessing a group of autonomous agents engaged in collective actions. Variables related to collective well-being are identified and measured, for instance resource availability, resource distribution, resource sustainability, or in general variables related to the welfare of the agents. These variables, specific to the problem at hand, will serve as the basis for calculating resilience. We measure these variables under normal conditions (reference curve) and during disruptions (performance curve). \textbf{The reference behavior is not necessarily an ideal behavior. Rather, the goal in this stage is to compare the system's behavior with and without disruptions.}

In the second stage, we recognize that systems often face successive adverse events over time. We assume that disruptions occur sequentially. To capture resilience at a specific moment, the system's response to all previous disruptions is analyzed within defined time windows, considering the timing of event occurrence, failure, and recovery. Resilience is then calculated within each window by comparing the performance and reference curves. In the third  stage, we aggregate the resilience metrics over time for each variable. This aggregation penalizes decreasing resilience and rewards improvement during the sequence of disruptions, reflecting the system's ability to learn from past disruptions. Finally in the last stage, the individual resilience measurement across all variables are then combined into a single resilience score.

Each stage is elaborated upon in detail as follows. Table \ref{tab:notationMethodology} includes some notation for better comprehension of this section.
\begin{table}[!h!t]

    \footnotesize
    \renewcommand*{\arraystretch}{1.1}
    \centering
    \caption{Symbol notation used in the methodology.}
    \label{tab:notationMethodology}
    \begin{tabular}{M{2cm}|J{8cm}} 
    \hline
    \textbf{Notation} & \multicolumn{1}{c}{\textbf{Description}} \\ \hline
    $N$ & Number of agents. \\ \hline
    $K$ & Number of variables measure for performance of the system. \\ \hline
    $L$ & Number of disruptive events in a time windows $[t_0, t_f]$. \\ \hline
    $P_{ij}(t)$ & Performance curve measure for agent $i$ and variable $j$ with $i \in 1 \leq i \leq N$ and $j \in 1 \leq j \leq K$. Measure in a time window $[t_0, t_f]$. \\ \hline
    
    $R_{ij}(t)$ & Reference curve measure for agent $i$ and variable $j$ with $i \in 1 \leq i \leq N$ and $j \in 1 \leq j \leq K$. Measure in a time window $[t_0, t_f]$. \\ \hline
    
    $P_{j}(t)$ & Performance curve assembly for all agents for variable $j$ with $j \in 1 \leq j \leq K$. \\ \hline
    
    $R_{j}(t)$ & Reference curve assembly for all agents for variable $j$ with $j \in 1 \leq j \leq K$.  \\ \hline

    $J_{jl}$ & Resilience measure for variable $j$ with $j \in 1 \leq j \leq K$, and disruptive event $l \in 1\leq l \leq L$.  \\ \hline

    $F_{jl}$ & Failure profile for variable $j$ with $j \in 1 \leq j \leq K$, and disruptive event $l \in 1\leq l \leq L$. \\ \hline

    $G_{jl}$ & Recovery profile for variable $j$ with $j \in 1 \leq j \leq K$, and disruptive event $l \in 1\leq l \leq L$. \\ \hline
    
    $J_{j}$ & Resilience measure for variable $j$ with $j \in 1 \leq j \leq K$ assembly in all disruptive event. \\ \hline

    $J$ & Resilience measure assembly for all indicator, and assembly for all disruptive event.\\ 
    
    \midrule
    \end{tabular}

\end{table}

\subsection{Stage I: Performance and reference Curves}
The objective of this stage is to define and measure variables related to collective well-being. Initially, we assume that there are $N$ interacting agents, depicted as circles in Fig. \ref{fig:methodologyResilience}. This stage entails defining $K$ variables related to collective well-being of the agents. For example, in an environment where the goal is resource consumption, variables could include resource availability, equality in access to resources, among others. Each variable is time dependent, with and without disruptive events, to establish performance and reference curves. These are shown in the magenta foreground of the second layer in Fig. \ref{fig:methodologyResilience}.

Performance curves for indicator $j$ and agent $i$ are denoted as $P_{ij}(t)$, with reference curves as $R_{ij}(t)$ (see \autoref{tab:notationMethodology}). The performance curves, initially gathered separately for each agent, are consolidated at this stage. Consolidation at the agent level is achieved using function $h$, producing a single curve $P_{j}(t)$ representing collective consolidation each indicator $j$. The same applies to reference curves. For example, to measure resource access equality, $h$ might use the Gini index to quantify equality based on resource consumption. The specific functional form of $h$ depends on the context and variable. Additionally, aggregated variables by agents can also be utilized. For instance, if measuring resource availability, this variable, which is part of the environment, is already aggregated by agents. In any case, at the end of this stage, performance and reference curves for each variable should be consolidated.

\subsection{Stage II: Computed summary metrics}

The purpose of this stage is to derive metrics that summarize resilient behavior for each adversity across time intervals and for each measured variable. The performance and reference curves are defined for the entire observation period $[t_0, t_f]$. Within this period, smaller time windows are used to isolate and analyze each disruptive event, as illustrated in the orange plane of Fig. \ref{fig:methodologyResilience}. Resilience metrics are calculated by comparing the performance curve with the reference curve for each time window. This process allows for detailed assessment of how the system behaves before, during, and after each disruption.

This summary value is denoted as $J_{jl}$ (variable $j$ and disruptive event $l$) and is calculated by the metric described in \cite{ayyub2014metrics}.  The metric involves identifying the failure profile, which relates to the speed and magnitude of the system's degraded behavior after the disruptive event.  Additionally, it also takes into account the recovery profile, considering the speed and stabilization of the system following the disruptive event.

The equation relating to the calculation of the summary metric denoted as $J_{jl}$ is as follows in Equation~\eqref{ec:summaryMetric}:
\begin{equation}
    J_{jl} = \dfrac{t_i + F_{jl}\Delta t_f + G_{jl} \Delta t_r}{ t_i + \Delta t_f + \Delta t_r},
    \label{ec:summaryMetric}
\end{equation}
where $F_{jl}$ corresponds to the failure profile, and $G_{jl}$ represents the recovery profile. $\Delta t_f$ and $\Delta t_r$ denote the failure and recovery event duration, computed as $\Delta t_f = t_f - t_i$ and $\Delta t_r = t_r - t_f$. The terms $t_i$, $t_f$, $t_r$ respectively represent the time of the incident occurrence, the time of failure when the performance deteriorates to the lowest point, and the recovery time, where it is assumed the system should reach a stable state. Total recovery is not necessarily expected, but the time $t_r$ is set as a reference to consider the recovery progress. 

The method for calculating the failure and recovery profiles is as follows:
\begin{align*}
    F_{jl} = \dfrac{\int_{t_i}^{t_f} P_{jl}(t)dt}{\int_{t_i}^{t_f} R_{jl}(t)dt},
\end{align*}
and
\begin{align*}
    G_{jl} = \dfrac{\int_{t_f}^{t_r} P_{jl}(t)dt}{\int_{t_f}^{t_r} R_{jl}(t)dt}.
\end{align*}

These profiles are positive measures. Values close to 1 indicate that observed and expected behaviors are nearly identical, suggesting minimal deviation from the reference interval. Values below 1 indicate performance below expectations, while values above 1 demonstrate behavior exceeding expectations for the performance curves. For this analysis to be meaningful, it is crucial that the well-being variables have a positive interpretation, meaning higher values correspond to better well-being.

At the conclusion of this stage, for each variable the summary metrics for all $L$ disruptive events are computed, resulting in a set of $J_{jl}$ index by variable and disruptive event.

\subsection{Stage III: Time-window assembly}
\label{subsec:timeWindowAssembly}

The proposed definition of resilience emphasizes transformation, suggesting that more resilient systems improve their behavior in response to disruptive events. Systems that adapt and learn from disruptions become better prepared for future occurrences, enhancing their ability to anticipate and respond to new events. Conversely, systems that fail to recover from disruptions may become more sensitive to future events, leading to decreased resilience. During this stage, efforts focus on penalizing behaviors where the system fails to transform between disruptive events, reducing resilience over time. Conversely, rewarding occurs when the system demonstrates improved resilience across events. Averaging across consecutive time-windows is proposed, penalizing decreases and incrementally rewarding increases. Therefore, the proposed metric rewards systems that not only recover and adapt but also show measurable improvements in well-being indicators, enhancing their ability to anticipate and manage future disruptions.

For each variable and across all disruptive events, a single metric $J_{j}$ (with $j \in 1 \leq j \leq K$ is computed. In the initial iteration of the calculation, it is performed as follows: 
\begin{align*}
         \left(\frac{J_{kl} + J_{k(l+1)}}{2}\right) \left( 1 + (J_{k(l+1)} - J_{kl})\right).
\end{align*}

The resulting metrics undergo an iterative process, consolidating into a single value. Negative variations indicate decreasing factors of the summary metric across successive disruptive events. If a negative variation results in a negative value, we apply saturation, forcing the value to zero. Conversely, positive variations, which indicate an increase in resilience through disruptive events, are saturated at 1 if they exceed this value.
This stage is represented in the green plane of Fig. \ref{fig:methodologyResilience}. At the conclusion of this stage, we obtain a measure $J_{k}$, which represents the resilience assembly for agents and disruptive events in a specific scenario.

\subsection{Stage IV: Variables assembly}

So far, we have the set $\{J_1, \cdots, J_K\}$ of summary metrics, one for each variable. However, coupling is necessary to generate a single metric across all $K$ variables. Typically, coupling summary metrics involves averaging or using a weighted average. However, since each indicator represents a component associated with well-being, it is expected that the coupling should penalize low values in the set, indicating poor performance in some variables. Therefore, the harmonic mean is proposed as the coupling metric. This stage is represented in the blue plane of Fig. \ref{fig:methodologyResilience}, where at the end of this stage, a single measure $J$ is obtained, representing the measure of resilience assembly among agents, disruptive events, and variable for well-being.

\section{Case Studies}
\label{sec:cases}

The objective of this section is to measure cooperative resilience in AI multiagent systems, and investigate how cooperative resilience manifests when these systems are subjected to disruptive events.  To achieve this, used Melting Pot 2.0 \cite{agapiou2022melting}, a research tool designed to study multi-agent AI systems. The specific scenario chosen is referred to as `Commons Harvest Open,' where multiple agents inhabit a confined space containing  trees laden with apples. The objective for each agent is to consume as many apples as possible. Consumed apples regenerate with a probability per step that depends on the number of remaining apples on the tree. If all the apples on a tree are consumed, the tree vanishes. In this scenario a social dilemma might arise, when all apples are depleted from a tree, no further apples with grow and this goes in detriment of the entire population. 

Currently, the primary metrics used to evaluate system performance in such scenarios focus on the number of resources consumed by the agents. However, this is insufficient to fully understand the dynamics at play, especially in a context where external disruptive conditions can significantly impact collective welfare. The social dilemma emulated in this scenario requires a deeper evaluation that considers not just resource consumption, but also how agents anticipate, prepare, resist, recover and transform from these disruptions. This underscores the need for a cooperative resilience metric that can assess how the system face of adversity.

To systematically assess resilience, we introduce two distinct disruptive events. As per the definition, events should pose a risk to the joint-welfare of the agents. The \textbf{first disruptive event} involves the sudden removal of apples from the environment, simulating resource depletion and testing the agents' ability to sustain the remaining trees. This event is characterized by the probability of occurrence($p_s$) and the severity of the depletion ($v_s$). The original environment in Melting Pot is modified to include the introduction of this event.

The second disruptive event is designed not to be contingent upon environmental conditions, but rather on the introduction of agents lacking established policies or decision-making methods. This event involves adding two bots that engage in unsustainable harvesting, symbolizing a breakdown in social behavior. The event is triggered at a specified time, with the duration of interaction varying across three experiments. This duration indirectly influences the magnitude of the disruption. By adjusting the timing and length of the bots' introduction, we can assess the system's cooperative resilience to internal disruptions caused by non-cooperative behaviors.

The decision-making of the agents is defined through two approaches: Reinforcement Learning and Large Language Models. In RL, agents are trained using Proximal Policy Optimization (PPO) algorithm \cite{schulman2017PPO}. Although the disruptive events are not explicitly included in the training, the agents may develop an implicit capacity to anticipate such events based on low apple counts observed during training episodes. On the other hand, for LLM-based agents, an adapter is developed to connect the Melting Pot environment with a language model. The model is informed about the environment through text descriptions and decides agent actions sequentially. In this approach, each agent makes and executes a plan before other agent to move. Once all agents have moved, it constitutes a complete round.  This contrasts with RL, where all agents move simultaneously at each time step. Besides, unlike RL, which relies on extensive pre-training, LLM support in pre-existing world knowledge embedded within the language model to reason and make decisions.  

\subsection{Reinforcement Learning-based agents}

The PPO algorithm \cite{schulman2017PPO}, a gradient-based optimization method aimed at maximizing the expected return of policies, is employed to train the agents using RL. The training is conducted using independent RL for each agent. This resulting in each agent possessing its own policy parameterized by its own neural network. The architecture of the network for each agent consisting of a feedforward neural network with two hidden layers, each with 64 neurons and ReLU activation function. This neural network is connected to another network that features a single hidden layer with 1280 neurons, also utilizing the ReLU activation function. 

The total training process span 1.280.000 steps, with mean duration episode 1500 steps. The training process took a total of 769 episodes. For training, the reward function of the problem is defined as 1 if an apple was eaten and 0 otherwise at each step. The agents are capable of executing the following actions: moving up, down, right, or left, rotating right or left, or shoot a laser beam that relocated agents within its range to a distant position from the apples.

\subsection{Large Language Model-augmented agents}
To work with an LLM in the decision-making process of each agent, we develop an adapter that converts the spatial observations received by each agent into textual observations that are comprehensible for a language model, specifically GPT-4. This process is divided into two parts: first, a converter is developed to transform the visual information into ASCII format, representing what each agent perceives about the environment at any given moment in real-time. Subsequently, this information is transformed into detailed textual descriptions of what is observed by each agent within their field of vision. 

\begin{figure}[htbp]
\centerline{ \includegraphics[width = 0.7\textwidth]{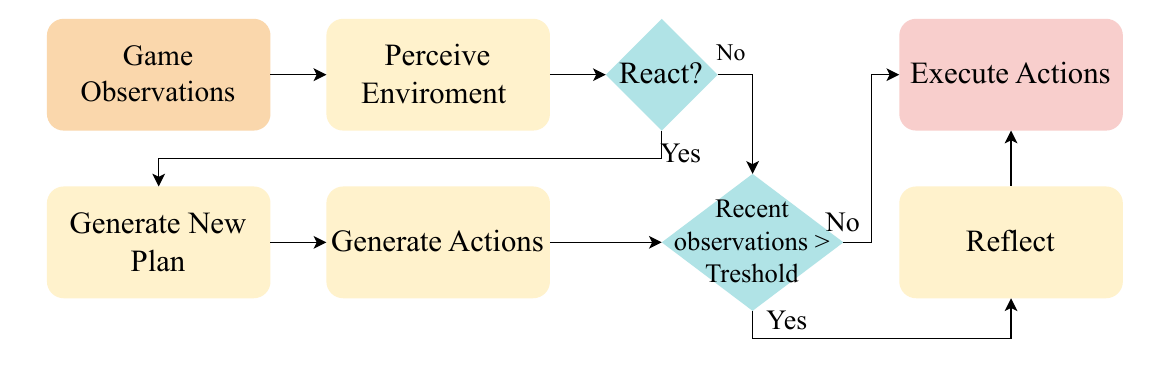}}
\caption{Diagram summarizing the reasoning process flow within the LLM architecture, leading to the action-taking phase of each agent. The diagram is inspired by the architecture proposed in \cite{mosquera2024LLM}.
}
\label{fig:architectureLLM}
\end{figure}

To integrate the adapter for working with an LLM in the decision-making process of each agent, the architecture ``Generative Agents'' proposal in \cite{mosquera2024LLM} is used. The architecture consists of a memory module, a perception module, a planning module, a reflection module, and an action module. Fig. \ref{fig:architectureLLM} shows a diagram of the modules comprising it. It is important to note that each module of the architecture is structured around a prompt directed at the language model, utilizing engineering practices of prompting that follow specific guidelines, similar to those applied in projects such as MetaGPT and AutoGPT. The prompts incorporate relevant details of each agent, adapting to current observations and pertinent memories. 

Furthermore, the ``Chain of Thought'' methodology \cite{prompting1}, except for the action module, is employed to structure the responses, enhancing the model's capability to reason and generate coherent outputs. Regarding the module responsible for action execution and decision-making within the simulated environment, an innovative prompting technique called ``SELF-DISCOVER,'' developed by Zhou et al. \cite{prompting2}, has been implemented. This technique enables the LLM to explore and apply complex reasoning structures autonomously enhancing its ability to generate adaptive responses. 

\subsection{Evaluation of cooperative resilience}

In the previously described environment, the phases of the methodology are followed to establish a value of cooperative resilience with both RL and LLM decision-making systems. The evaluation is conducted for a set of experiments proposed in the two disruptive events. Below are the detailed parameters of the experiments conducted.

\subsubsection{First type of disruptive events: apple disappearance}
To cover various scenarios related to the probability of occurrence of a disruptive event and its probabilistic magnitude, the probability of the disruptive event occurring is fixed at certain points in the simulation, with $p_s = 1$ at specific moments and $p_s = 0$ at others. For the impact level, the event iterates through each apple and removes this with a probability corresponding to $v_s$, ensuring that at least one apple remains on each tree. Three probabilistic values for $v_s$ are considered, leading to nine scenarios described in the \autoref{tab:experiments}, the cells in the table with darker hues indicating higher disruption.
\begin{table}[!h!t]
    \footnotesize
    \renewcommand*{\arraystretch}{1.1}
    \centering
    \caption{Characterization of experiments with first disruptive event.}
  
    \label{tab:experiments}
    \begin{tabular}{|>{\columncolor{gray!30}}M{1.3cm}|J{1.8cm}|J{2.1cm}|J{1.9cm}|}
    \hline 
  
   \textbf{Time-step / round} & \cellcolor{myblue!10} \textbf{$v_s =$ 0.3} & \cellcolor{myblue!18} \textbf{$v_s =$0.5} & \cellcolor{myblue!25} \textbf{$v_s =$0.7} \\ \hline
    
    \cellcolor{gray!5} [250] / [25] & \cellcolor{level1!15}\textbf{E1:} one disruption and low magnitude.  &  \cellcolor{level2!15} \textbf{E2:} one disruption and medium magnitude. & \cellcolor{level3!15} \textbf{E3:} one disruption and highest magnitude. \\ \hline
    
    \cellcolor{gray!25} [50, 250] / [5, 25] & \cellcolor{level1!25} \textbf{E4:} two disruptions and lowest magnitude. & \cellcolor{level2!25} \textbf{E5:} two disruptions and medium magnitude. & \cellcolor{level3!25} \textbf{E6:} two disruptions and high magnitude. \\ \hline

     \cellcolor{gray!45} [50, 250, 400] / [5, 25, 40] & \cellcolor{level1!35} \textbf{E7:} three disruptions and low magnitude. & \cellcolor{level2!35} \textbf{E8:} three disruptions and medium magnitude. & \cellcolor{level3!35} \textbf{E9:} three disruptions and high magnitude.  \\ \hline 

    \end{tabular}

\end{table}

In the initial phase of the methodology, within the described environment, we have decided to incorporate curves related to resource availability, sustainability, and distribution. These factors are essential components of collective well-being. These dimensions are explored through the following specific measures: \textbf{(1)} apples alive \textit{per capita}, \textbf{(2)} trees alive \textit{per capita}, \textbf{(3)} cumulative gini equality index and \textbf{(4)} collective hunger level Index\footnote{Detailed descriptions of these indicators, including their theoretical foundations are provided in supplementary file.}.   

Fig. \ref{fig:somePerformanceCurves} shows some examples of the performance and references curves taken in the initial phase of the methodology. In Fig. \ref{fig:somePerformanceCurves}, it is observed that a disruptive event significantly influences the system across four key metrics. A direct impact is seen on the metric related to the number of apples in the environment, while a decline in expected performance is also noted in the other metrics. In the case of the number of living trees, the reference curve is situated above the value for the scenario presented. A similar trend is noted for the cumulative gini equality index, where, although the effect is not immediate, the disruptive event, over time, leads to a degradation in performance in comparison to the reference scenario. Regarding the hunger index for the depicted scenario, a significant disruption is not immediately evident. However, as events unfold, an increase in `hunger' becomes apparent, which makes sense given the scarcity of resources. Another important aspect is that, in the case of LLM decision-making techniques, unlike RL, behavior is not optimized according to resource availability. This results in rapid apple harvesting and eventual total tree disappearance due to disruptive events. These elements are analyzed for the experiments presented. However, depending on the experiment, different behaviors may be observed. 
\begin{figure}[h!]
\footnotesize
     \centering
     \begin{subfigure}[b]{0.24\textwidth}
         \caption{}
         \centering
        \includegraphics[width=\textwidth]{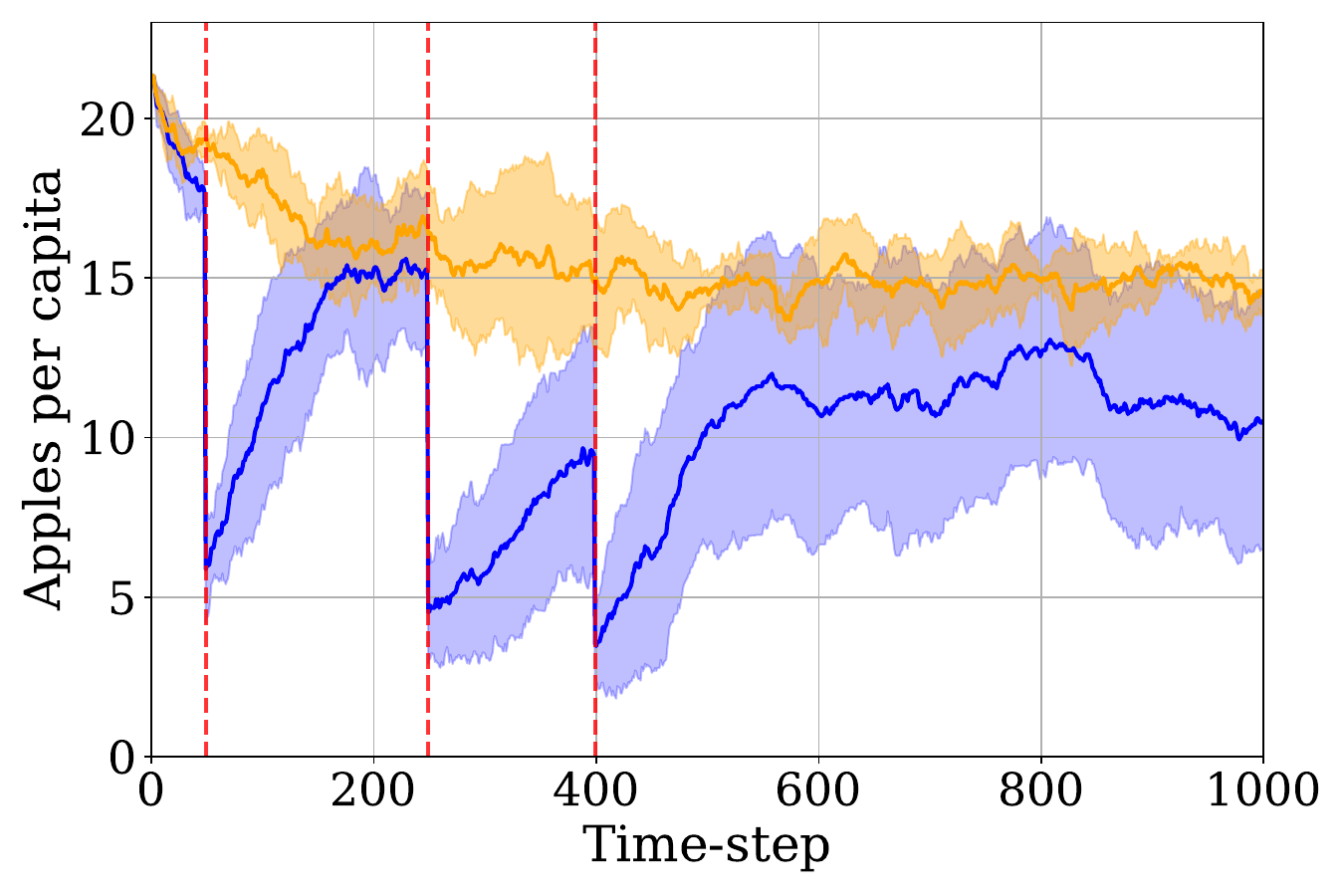}
         \label{fig:applesPerCapitaExample}
     \end{subfigure}
     \begin{subfigure}[b]{0.24\textwidth}
          \caption{}
        \includegraphics[width=\textwidth]{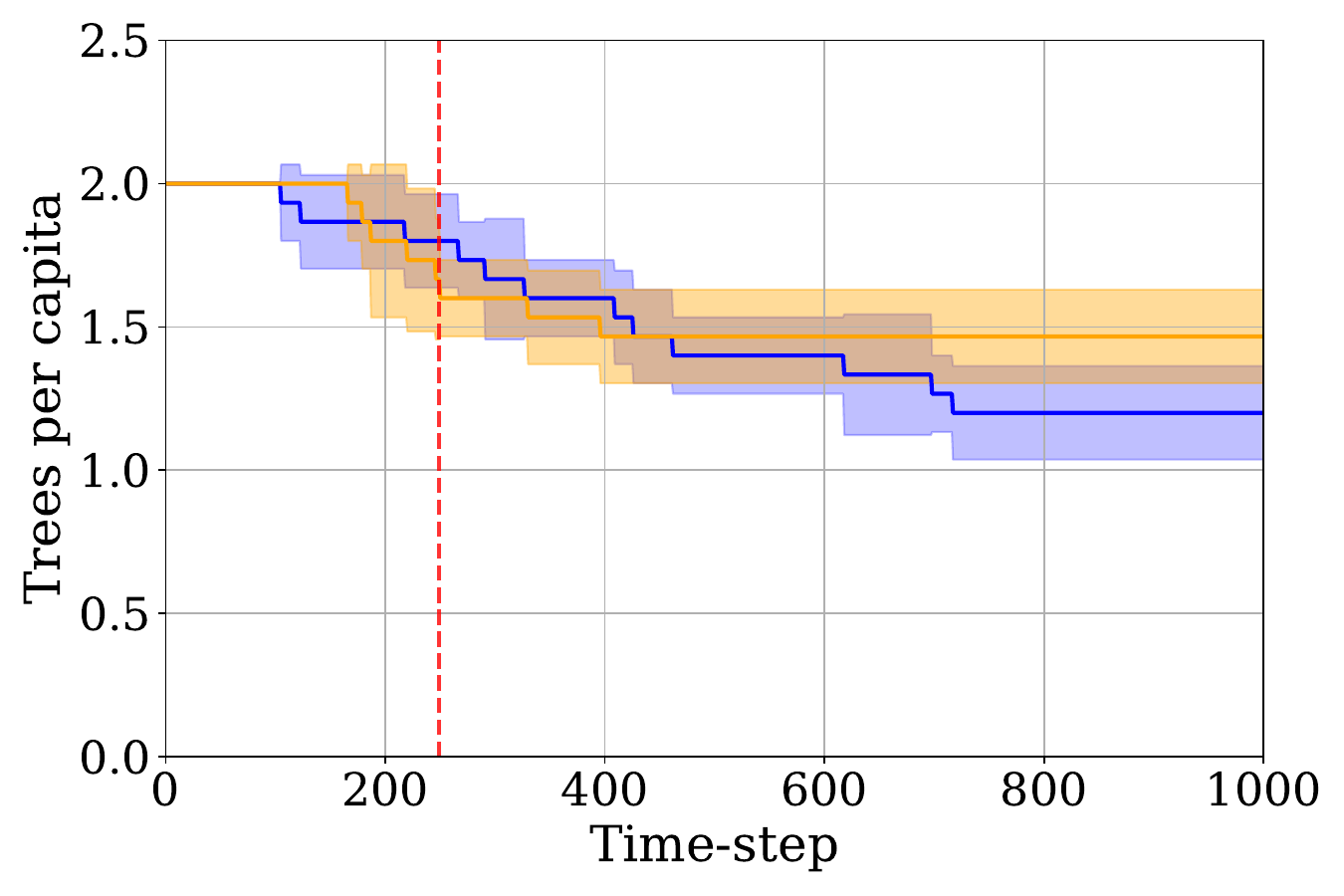}
         \label{fig:treesPerCapitaExample}
     \end{subfigure}
     \begin{subfigure}[b]{0.24\textwidth}
         \caption{}
        \includegraphics[width=\textwidth]{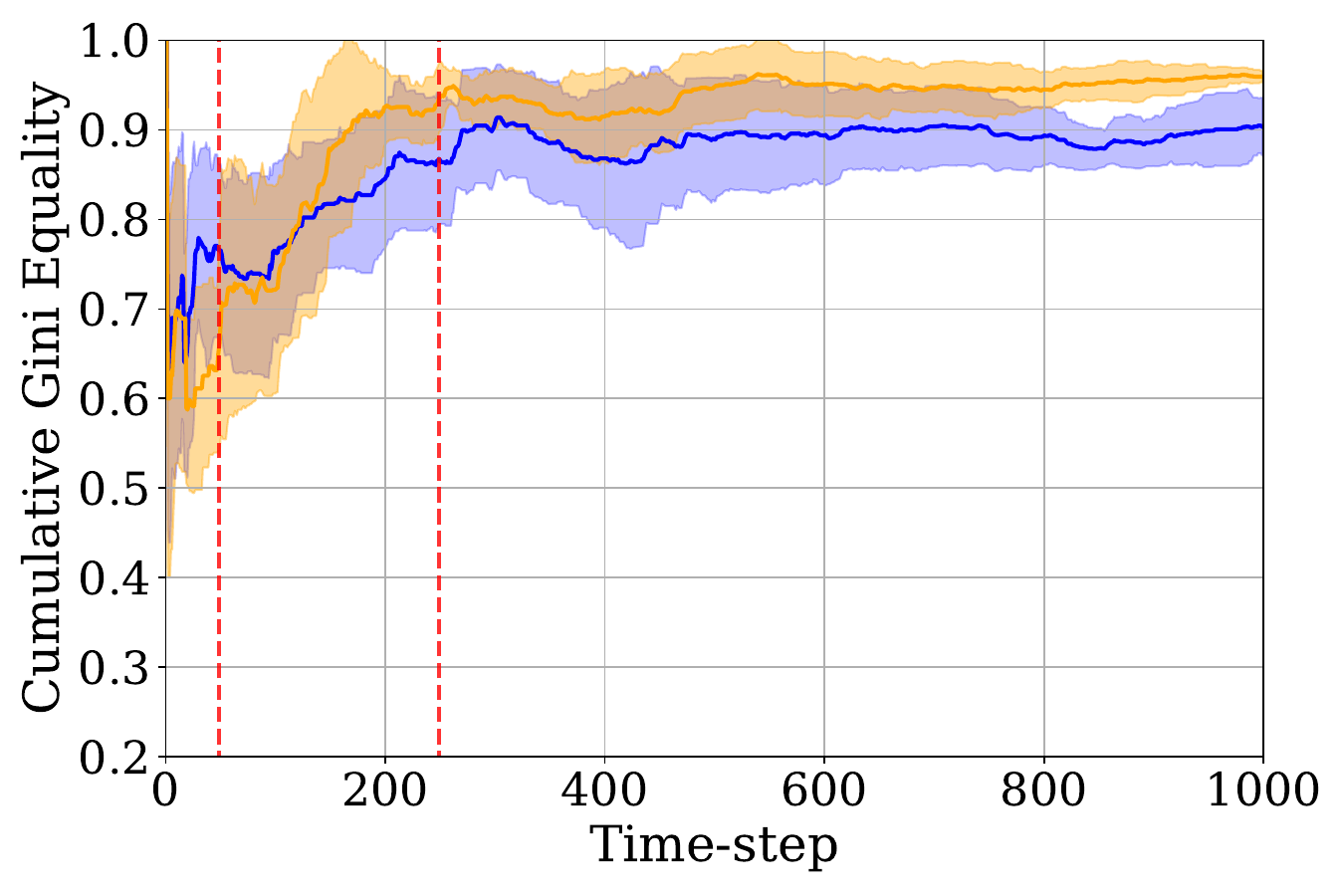}
         \label{fig:giniExample}
     \end{subfigure}
     \begin{subfigure}[b]{0.24\textwidth}
         \centering
         \caption{}
         \includegraphics[width=\textwidth]{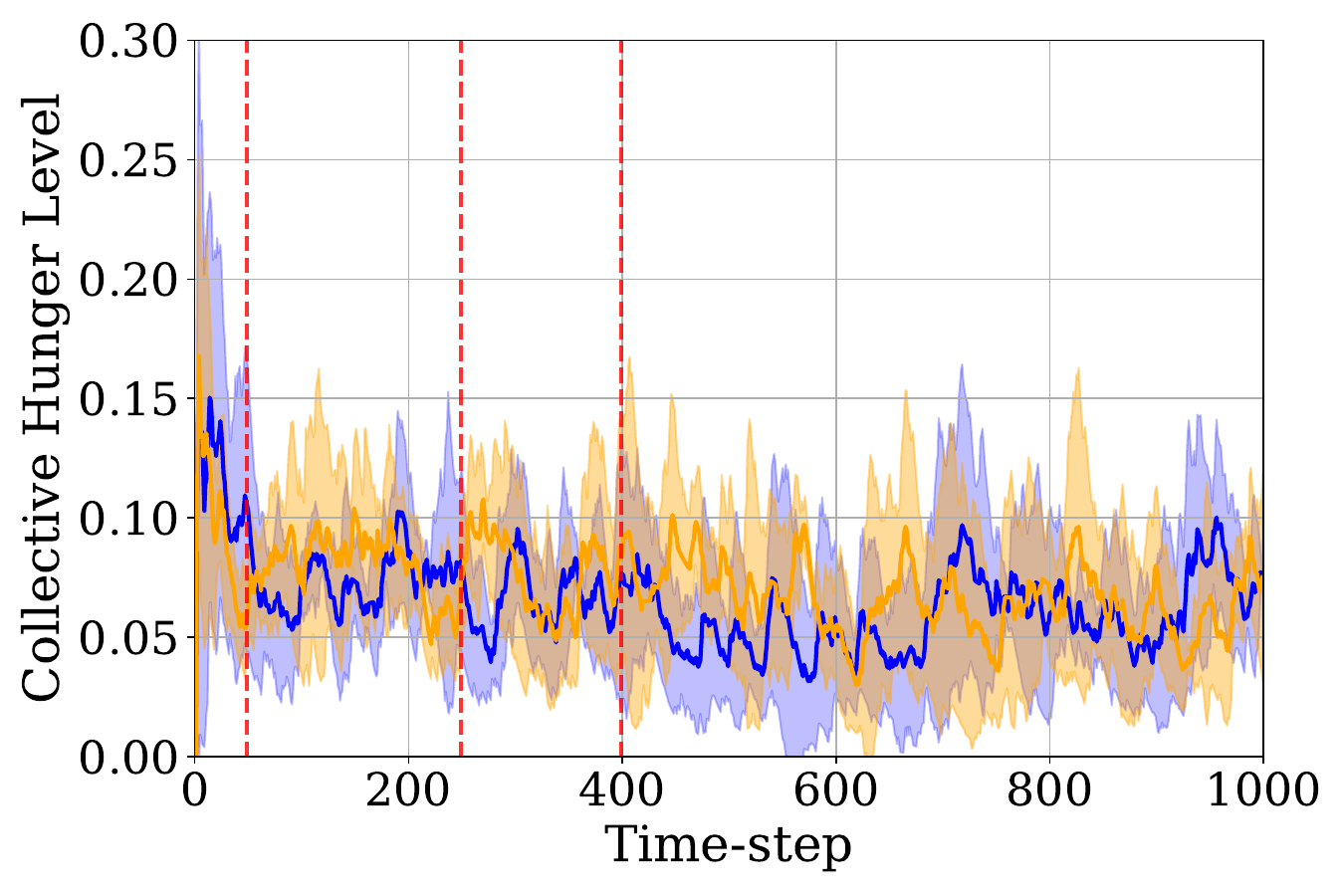}
         \label{fig:hungryExample}
     \end{subfigure}\\
     
     \begin{subfigure}[b]{0.24\textwidth}
         \centering
         \caption{}
        \includegraphics[width=\textwidth]{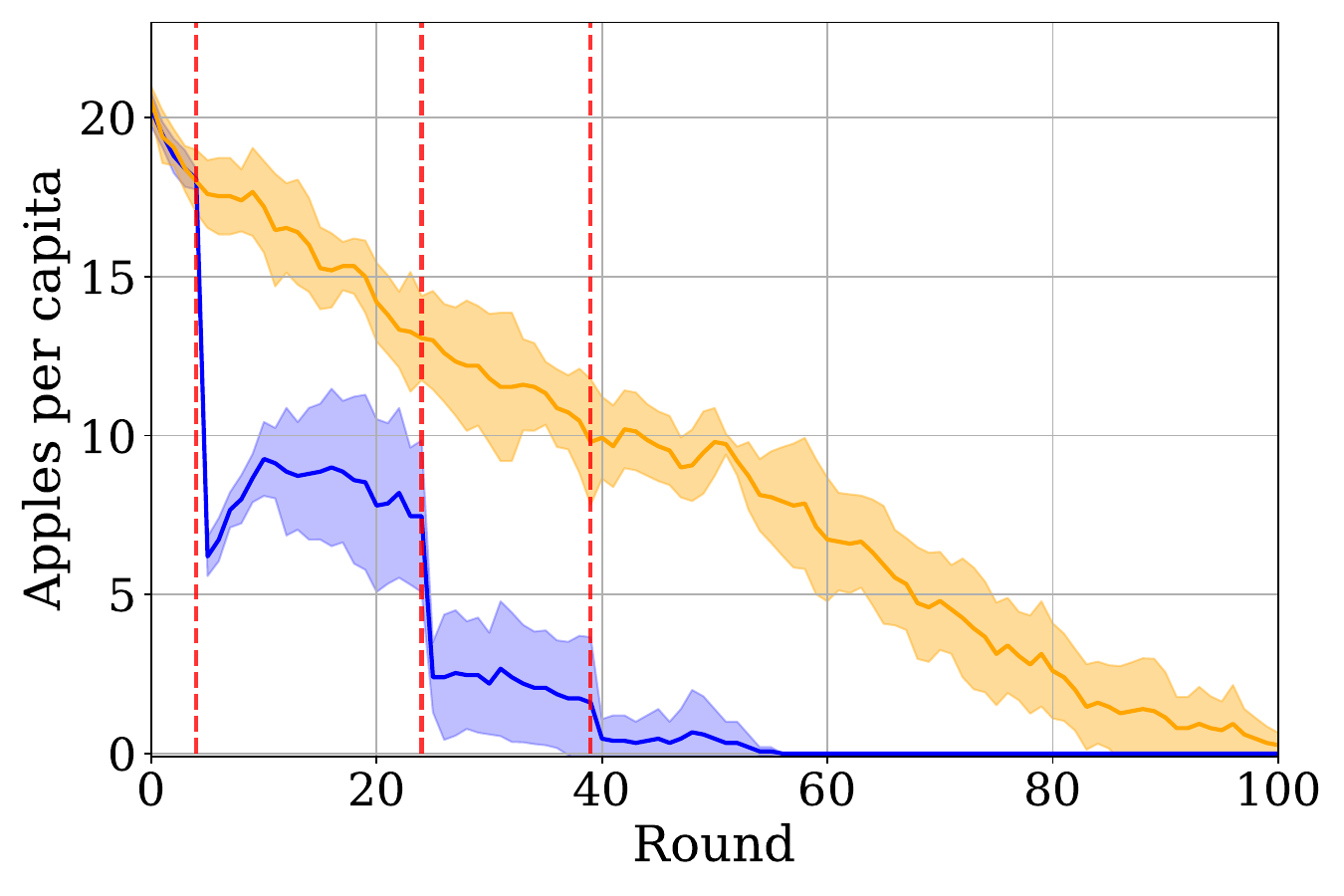}
        \label{fig:applesPerCapitaExampleLLM}
     \end{subfigure} 
     \begin{subfigure}[b]{0.24\textwidth}
         \centering
         \caption{}
         \includegraphics[width=\textwidth]{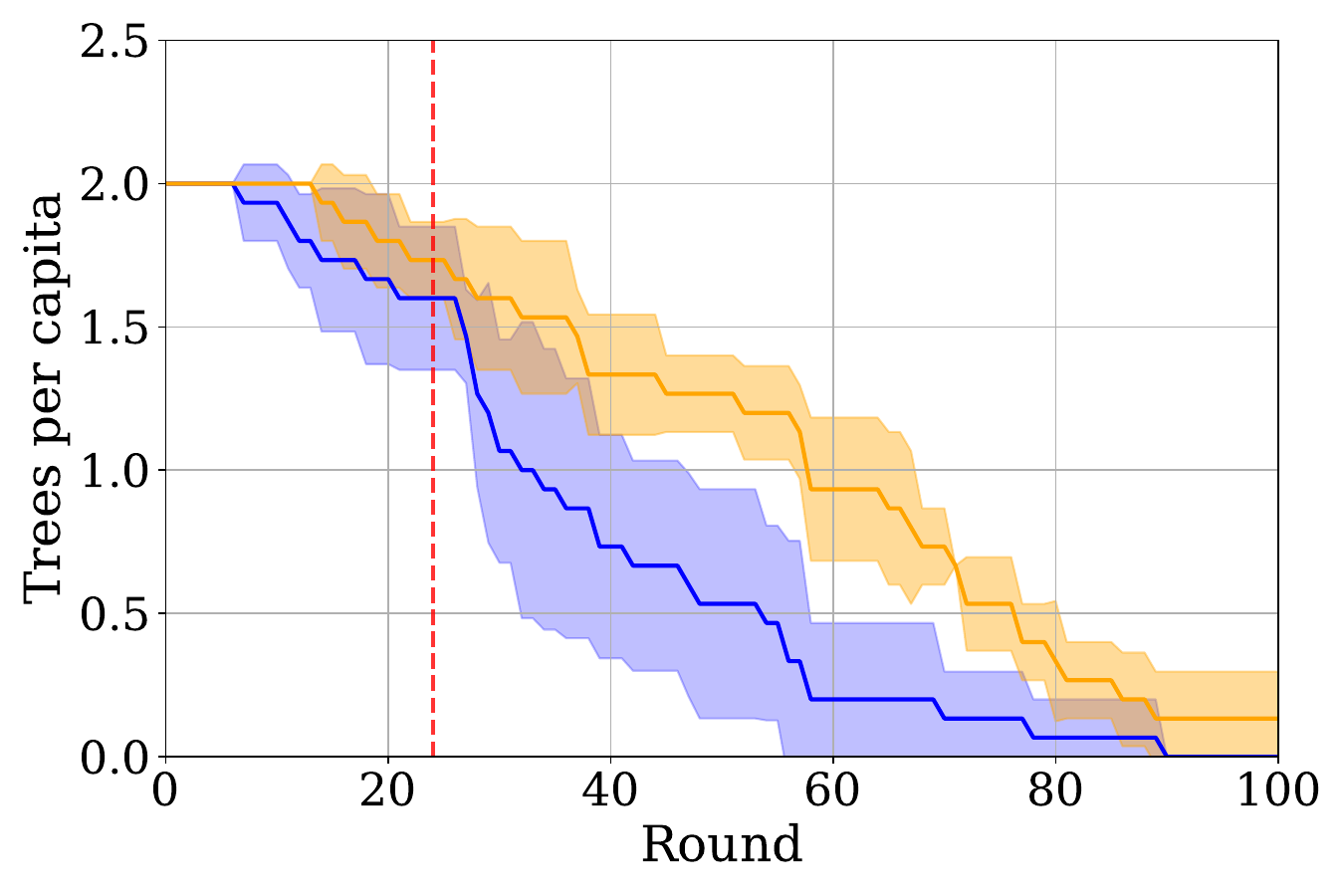}
    \label{fig:treesPerCapitaExampleLLM}
     \end{subfigure} 
     \begin{subfigure}[b]{0.24\textwidth}
         \centering
         \caption{}
        \includegraphics[width=\textwidth]{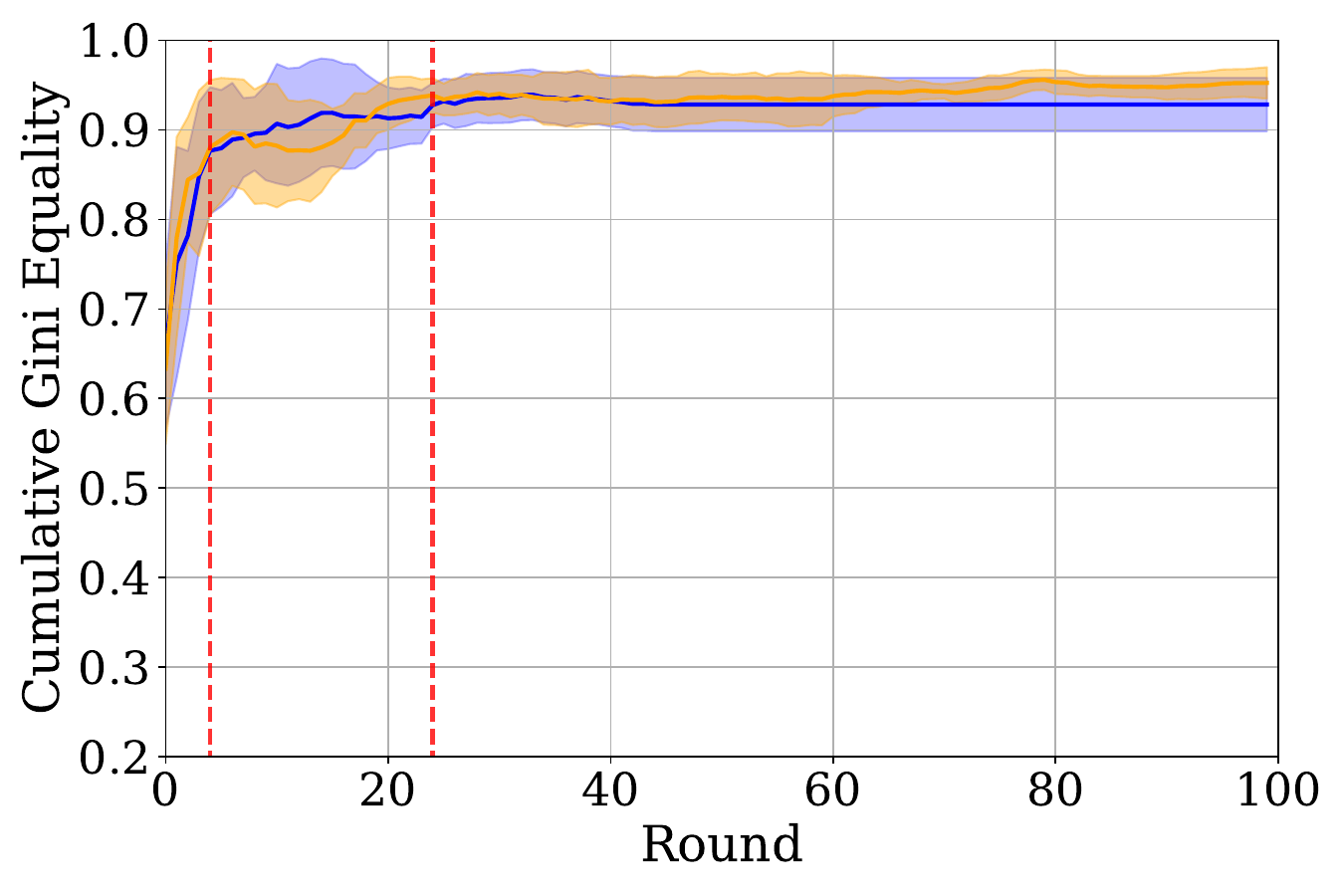}
         \label{fig:giniExampleLLM}
     \end{subfigure}
     \begin{subfigure}[b]{0.24\textwidth}
         \centering
         \caption{}
         \includegraphics[width=\textwidth]{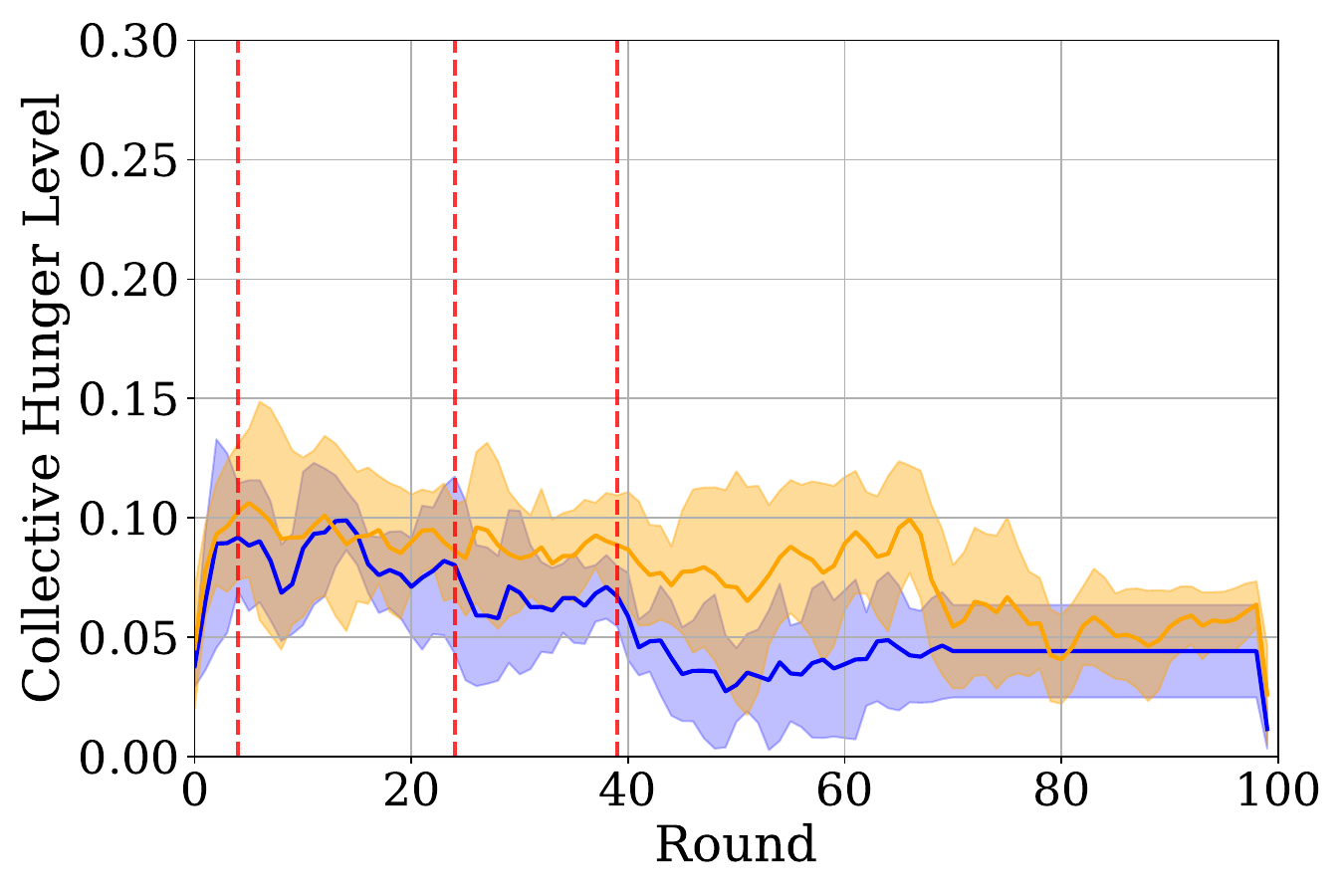}
         \label{fig:hungryExampleLLM}
     \end{subfigure}
   
    \caption{Performance and reference curves: The blue line represents the mean performance curve over five episodes, while the orange line indicates the mean reference curve. The shaded regions correspond to the standard deviation. The red dashed line marks the occurrence of the disruptive event. The top row (a, b, c, d) shows the results of agents trained with RL, while the bottom row (e, f, g, h) displays the results of LLM-based models. (a) and (e) depict the apples alive \textit{per capita} in experiment E9. (b) and (f) show the trees alive \textit{per capita} in experiment E2. (c) and (g) illustrate the Gini Equality Index in experiment E5. Finally, (d) and (h) present the Collective Hunger Level in experiment E7.}
     \label{fig:somePerformanceCurves}
\end{figure}

\begin{figure}[ht!]
\centering
\begin{subfigure}[b]{0.4\textwidth}
         \centering
        \caption{}
         \label{fig:resilienceRLMap}\includegraphics[width=\textwidth]{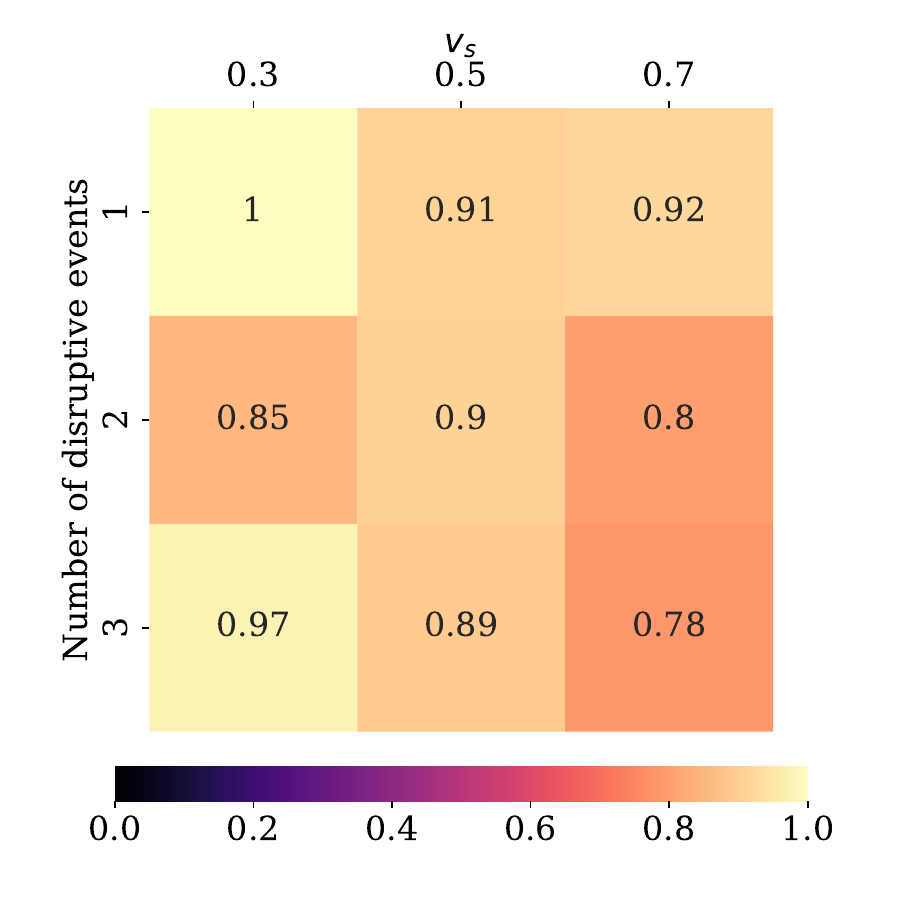}
         
     \end{subfigure}
     \begin{subfigure}[b]{0.4\textwidth}
         \centering
        \caption{}
         \label{fig:resilienceLLMMap}\includegraphics[width=\textwidth]{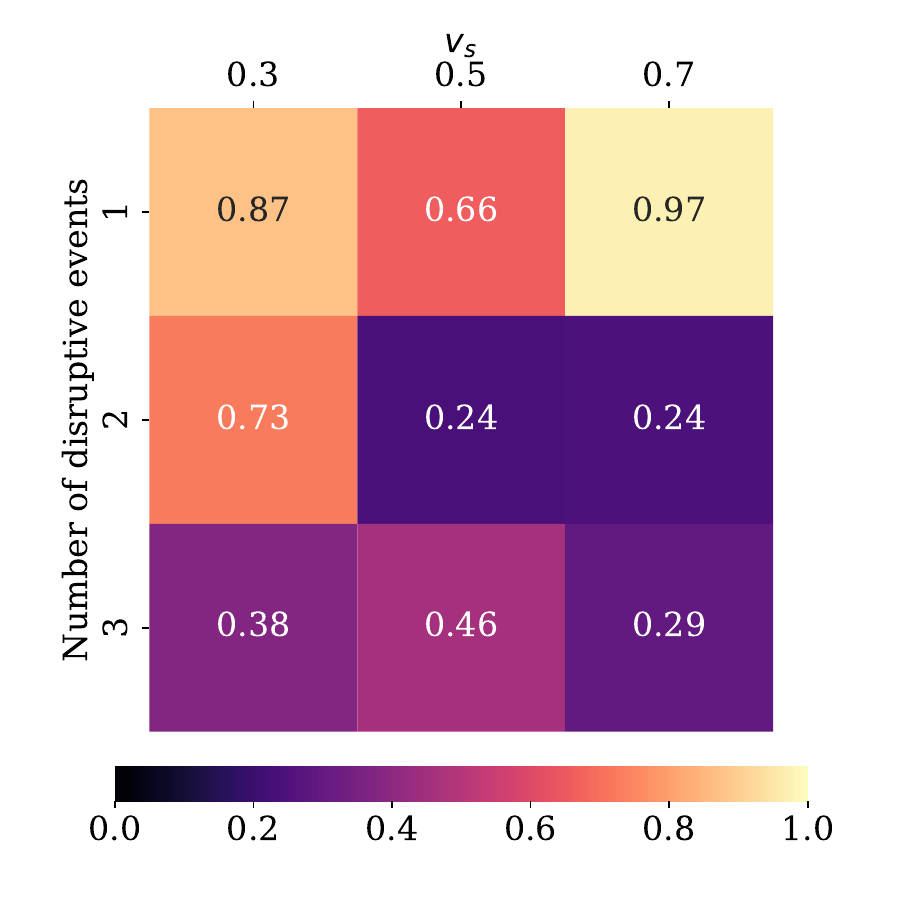}
         
     \end{subfigure}

\caption{Cooperative Resilience Map: This heatmap illustrates the impact of varying the number of disruptive events (1, 2, or 3) and the disturbance magnitude ($v_s$) on system resilience. The map uses darker colors to represent lower resilience values. Figure (a) shows results for the RL approach, while (b) displays results for the LLM.}
\label{fig:resilienceMap}
\end{figure}

After deriving the curves, the methodology conducts an ensemble analysis over time-windows. Critical milestones: incidence time, failure time, and recovery time are identified within the windows where the disruptive event occurs. The failure time marks the system's lowest performance level between the incidence and recovery times. Using these milestones, a summary metric is calculated, followed by ensemble analysis to produce a single resilience value for each variable. Then, an ensemble analysis is performed to obtain a final value across all variables. Fig. \ref{fig:resilienceMap} presents a heat map showing the cooperative resilience values in the proposed scenarios.

The cooperative resilience map in Fig. \ref{fig:resilienceMap} reveals a general trend: resilience decreases as the disturbance magnitude and the number of disruptive events increase, with this effect being more pronounced in LLM than in RL. Interestingly, in the RL approach, when $v_s = 0.3$, resilience is higher with three disruptive events (E7) compared to two (E4). This counterintuitive result can be attributed to the \textbf{transformation property} of the proposed methodology, which suggests that the system may be rewarded for improving its tolerance to successive disruptions after experiencing previous ones. This phenomenon underscores the importance of using resilience as a metric, as it captures dynamic system behaviors and adaptive capacities. The previous situation is also replicated in experiments E5 and E8 of LLM. In this case, we would expect resilience to reach a higher value for experiment E5 compared to E8. However, this is not the case. This outcome can be explained similarly to what was observed in RL.

Contrary to expectations, where a higher impact represented by $v_s$ would typically result in greater system degradation, the results show a lower resilience value in RL for the scenario with two events and $v_s = 0.3$ (E4) compared to $v_s = 0.5$ (E5). Similarly, in LLM, both $v_s = 0.5$ (E5) and $v_s = 0.7$ (E6) with two disruptive events yield the same resilience values. This deviation from the hypothesis suggests that a higher magnitude does not necessarily lead to poorer system recovery. Factors such as agent dynamics, environmental conditions, and scenario variability likely contribute to this unexpected behavior.

This highlights the importance of the cooperative resilience metric, as it captures complex interactions, adaptive responses and  capacities that conventional metrics may overlook, providing a more understanding of the system's performance under disruptions. Additionally, the cooperative resilience metric allows for a comparison between the two decision-making systems, RL and LLM. RL models generally outperform LLM, owing to their training on diverse episodes with varying apple values. In contrast, LLM, which lacks a training phase and generates actions based on its architecture alone, demonstrates an incomplete understanding of how actions impact resource depletion. This indicates potential advantages of employing cooperative-focused LLM architectures.

\subsubsection{Second type of disruptive events: unsustainable bots}

In the second disruptive event, unsustainable bots are introduced into the simulation. These bots are introduced at a standardized point: the 10th round in LLM or the 100th time-step in RL. The impact of this disruption is evaluated based on the duration of the bots' interaction with the environment. Three levels of disruption are considered: \textbf{E1:} Bots interact for 25 time-steps in RL and 5 rounds in LLM. \textbf{E2:} Bots interact for 50 time-steps in RL and 10 rounds in LLM. \textbf{E3:} Bots interact for 75 time-steps in RL and 15 rounds in LLM.
  
These duration intervals are chosen to ensure a proportionate comparison of the disruption impact across both approaches. In the case of LLM-agumented agents, one round consists of all the steps in which agents execute their plan in turns. The bots were configured to move only half the time within a round, meaning they do not move every step but alternate instead. The evaluation of cooperative resilience employs the same variables as the previous disruptive event. 

\begin{figure}[ht!]
\centering
\begin{subfigure}[b]{0.4\textwidth}
        \centering
        \caption{}
        \includegraphics[width=0.9\textwidth]{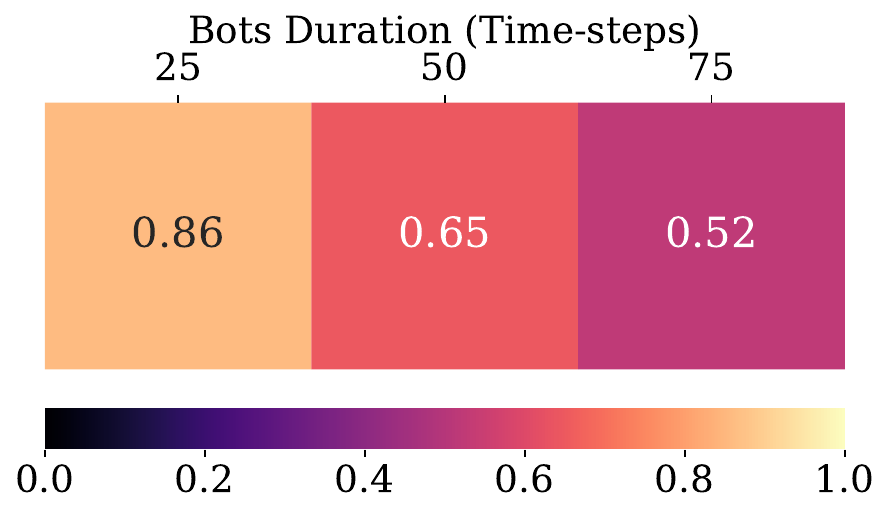}
         
         \label{fig:resilienceRLMapBots}
     \end{subfigure}
     \begin{subfigure}[b]{0.4\textwidth}
         \centering
        \caption{}
        \includegraphics[width=0.9\textwidth]{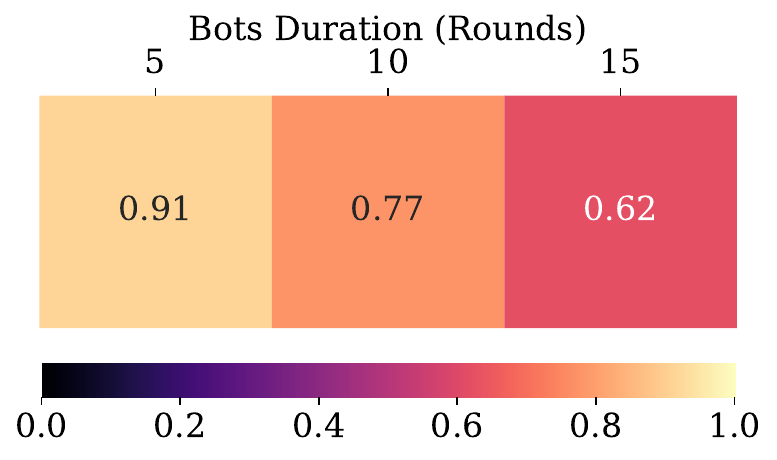}
         
         \label{fig:resilienceLLMMapBots}
     \end{subfigure}

\caption{Cooperative resilience map: This heatmap illustrates the impact of varying bot interaction durations on system resilience. Darker hues represent lower resilience values. Figure (a) shows the results for the RL approach, while (b) displays the results for the LLM-augmented agents.}
\label{fig:resilienceMapBots}
\end{figure}

Fig. \ref{fig:resilienceMapBots} presents the cooperative resilience measure across the three experiments. The results demonstrate that as the duration of bot interaction in the simulation increases, resilience values decrease. This outcome is anticipated, given that the bots consume resources unsustainably, directly and indirectly impacting the variables considered for resilience measurement.
Resource scarcity directly affects the availability of apples in the environment and the survival of trees. Moreover, bots that consume large quantities of apples create inequality in resource access, a situation that is exacerbated with prolonged bot presence.

\begin{figure}[ht!]
     \centering
     \begin{subfigure}[b]{0.48\textwidth}
         \centering
        \caption{}
        \includegraphics[width=0.8\textwidth]{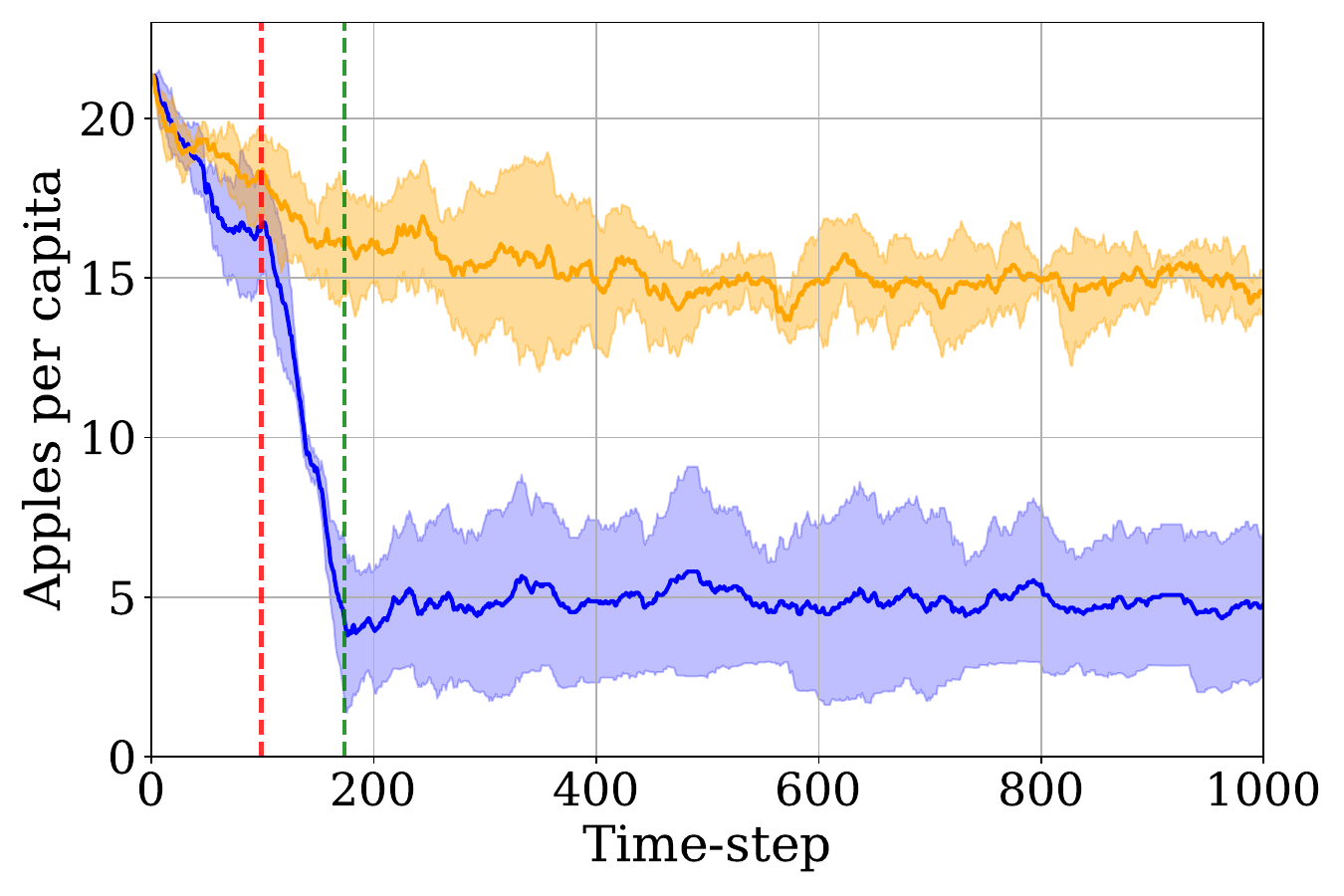}
         
         \label{fig:applesExamplesE2_secondDisruptive}
     \end{subfigure}
     \begin{subfigure}[b]{0.48\textwidth}
         \centering
         \caption{}
         \includegraphics[width=0.8\textwidth]{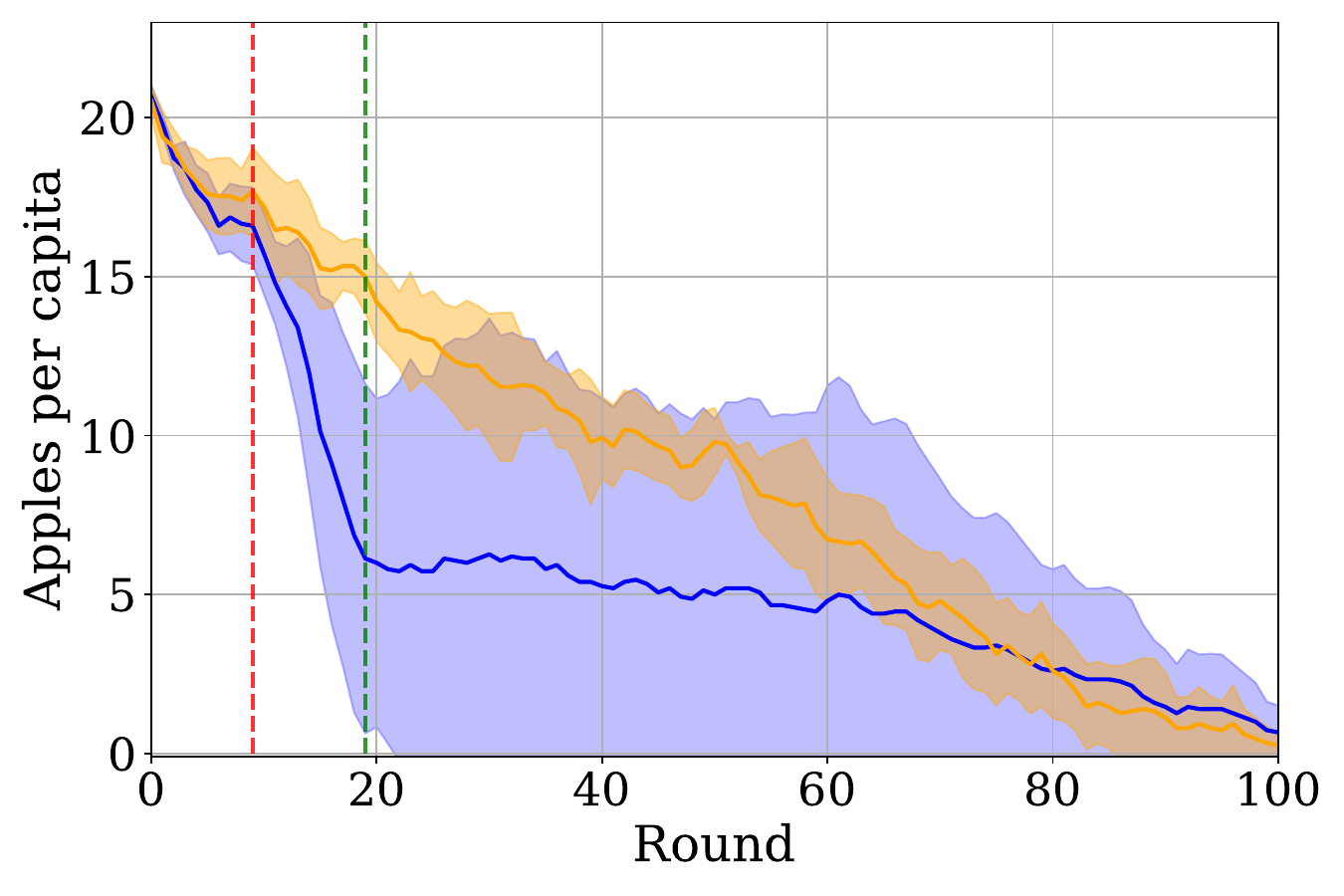}
         
         \label{fig:applesExamplesE2LLM_secondDisruptive}
     \end{subfigure} 
   
    \caption{Apples alive \textit{per capita}. The blue line is the mean value throughout 5 episodes of the performance curve and the orange line is the mean of reference curve. The shaded areas indicate the standard deviation. The red dashed line shows the occurrence of the disruptive event. (a) E3 in RL and (b) E2 in LLM.}
     \label{fig:applesSecond}
\end{figure}

When comparing the results between RL and LLM techniques, it is evident that the values for LLM surpass those of RL across all three experiments. This suggests that, for this disruptive event, the LLM technique exhibits better recovery and failure profiles. Expanding on the previous observation, certain indicators in RL suggest that agents continue their resource consumption policy regardless of the disruptive event. For instance, in Fig. \ref{fig:applesExamplesE2_secondDisruptive}, the slope of apple availability for agents remains unchanged post-bot intrusion. This implies that the bots' presence does not alter resource consumption patterns once they depart. The previous situation differs in LLM. For instance, in Fig. \ref{fig:applesExamplesE2LLM_secondDisruptive}, once the bots leave, the slope changes, promoting a return to the expected behavior. This indicates that agents in the LLM framework may adapt their strategies based on the bots' actions, leading to a more socially adaptive behavior.  

This adaptability observed in LLM is captured effectively by our cooperative resilience metric, underscoring its significance. The metric not only reveals how systems recover from disruptions but also highlights how agents adjust their behaviors in response to external social influences. This ability to measure and analyze adaptive responses demonstrating why our cooperative resilience metric is a valuable tool in understanding the system's robustness to disruptions.

\subsection{Discusion of the Results}
The results from both types of disruptive events highlight the complex dynamics that influence the system's resilience. These findings underscore the need for a broader understanding of how varying magnitudes and frequencies of disruptive events affect the final value of cooperative resilience. While additional experiments could enhance the accuracy of the cooperative resilience metric, the current methodology effectively captures resilience by focusing on the process of confronting disruptions, an aspect often overlooked by traditional metrics in AI multiagent systems. This approach is particularly valuable for studying AI systems prone to failures, as it addresses the impact of disruptions and highlights the agents' adaptive responses. Despite these complexities, the methodology can be appropriately applied to these experiments, establishing a cooperative resilience value that aligns with the definition outlined in this document.

\section{Conclusion and Future Work}
\label{sec:conclusions}

In this article, we have introduced the concept of cooperative resilience in cooperative AI, a notion proposed after analyzing the definition of resilience across various domains and aligning it with the concepts and scope of cooperative AI. This contribution not only try to unify the terminology within cooperative AI, but also aligns with interdisciplinary research efforts to understand emergent resilience in complex systems. Following the establishment of this definition, we have proposed a methodology designed to quantify cooperative resilience consistently with the defined concept. This methodology aims to estimate resilience in cooperative AI systems with the aim of enhance or comparing this value in future studies. 

The proposed methodology is applied to and validated in experiments using Melting Pot 2.0, specifically the `Common Harvest Open' scenario, employing both RL and LLM approaches for the control of agents. Two sets of experiments were conducted, one involving a disruptive event related to vanished apples at some point, and the other involving a disruptive event associated with the inclusion of bots. Under varying conditions of disruptive events, the value of cooperative resilience is determined in 9 experiments in the first case and in 3 experiments in the other disruptive event. The interplay of factors contributing to system resilience is underscored by the results, revealing instances where unexpected recovery patterns are demonstrated by the systems. Notably, the observation that resilience can sometimes increase with the number of disruptive events or vary with the magnitude of disruption, challenges conventional ideas and suggests a complex adaptive capacity inherent in these systems. These results, while preliminary, provide a foundation for deeper investigation into the dynamics of cooperative resilience and highlight the need for a broader range of experiments to understand the behavior of resilience.

Moreover, this research opens avenues for interdisciplinary collaboration, drawing parallels with resilience studies in ecology, psychology, network science, and other domains. Such collaborations can enrich our understanding of resilience as a multi-faceted concept and foster the development of more resilient cooperative AI systems. 

Future research should aim to expand the experimental framework to encompass a broader range of scenarios and disruptive events. Applying the developed methodology to experiments involving human performance could also enable comparisons between machine-only decision-making and human-machine interactions, providing valuable insights into cooperative resilience. Furthermore, a deeper exploration of the factors contributing to the emergence of resilience would be beneficial. Inverse problem approaches, such as inverse games and inverse reinforcement learning, can help uncover the underlying motivations driving resilient behaviors, facilitating their replication and enhancing resilience properties in AI systems.

\bibliographystyle{IEEEtran}
\bibliography{references}

\end{document}